\documentclass{article}

\newcommand{\name}{MASTER}

\usepackage[normalem]{ulem}
\usepackage{soul}
\usepackage{float}
\restylefloat{table}
\usepackage{graphicx}
\usepackage{siunitx}
\usepackage[left=2cm,right=2cm]{geometry}

\begin{document}

\title{
 A Machine Learning Accelerator In-Memory for Energy Harvesting
}

\author{
  Salonik Resch\\
  \texttt{resc0059@umn.edu}
  \and
  S. Karen Khatamifard \\
  \texttt{khatami@umn.edu}
  \and
  Zamshed Iqbal Chowdhury \\
  \texttt{chowh005@umn.edu}
  \and
  Masoud Zabihi \\
  \texttt{zabih003@umn.edu}
  \and
  Zhengyang Zhao \\
  \texttt{zhaox526@umn.edu}
  \and
  Jian-Ping Wang \\
  \texttt{jpwang@umn.edu}
  \and
  Sachin S. Sapatnekar \\
  \texttt{sachin@umn.edu}
  \and
  Ulya R. Karpuzcu \\
  \texttt{ukarpuzc@umn.edu}
}

\date{}
\maketitle

\thispagestyle{empty}

\begin{abstract}
\vspace{.1cm}
\noindent 
There is increasing demand to bring  machine learning capabilities to low power devices. By integrating the computational power of machine learning with the deployment capabilities of low power devices, a number of new applications become possible. In some applications, such devices will not even have a battery, and must rely solely on energy harvesting techniques. This puts extreme constraints on the hardware, which must be energy efficient and capable of tolerating interruptions due to power outages. Here, as a representative example, we propose an in-memory support vector machine learning accelerator utilizing non-volatile spintronic memory. The combination of processing-in-memory and non-volatility provides a key advantage in that progress is effectively saved after every operation. This enables instant shut down and restart capabilities with minimal overhead. Additionally, the operations are highly energy efficient leading to low power consumption.
\end{abstract}

\section{Introduction}
Machine learning is desirable for low-power, edge  devices as it provides the capability to solve a wide variety of problems. As a result, much research has been devoted to optimizing hardware for machine learning inference on such devices \cite{conti2018xnor,pudiannao}. Going even further, energy harvesting techniques \cite{energyharvestingkim2014ambient} remove the need for a battery, enabling the placement of such devices into almost any conceivable environment. There are many exciting possible applications, such as low power sensor networks \cite{sensornetworkmanic2016intelligent}, wearable tech, or even implants \cite{implantgreenspan2016guest}. Previous work has already experimentally demonstrated machine learning capability on energy harvesting devices using commercially available hardware \cite{InferenceBeyondEdge}.  

Energy harvesting applications present numerous and unique challenges. Power limitations are extreme. The energy harvested from the environment is likely far less than what can be supplied by a battery. Thus, energy efficiency is even more critical than in mobile applications. Significantly, the process of energy harvesting also introduces the requirement for \textit{intermittent processing}. Energy sources (such as sunlight, heat, movement) may be unreliable, and a device will have to shut down when the power source goes away. Additionally, even when available, the power source may be insufficient to run the device continually. In order to operate within the power budget, the device must acquire energy over time and consume it in bursts \cite{chargeusechandrakasan2008next}. Intermittent processing introduces new considerations and metrics for performance \cite{processor}. Significantly, correctness has to be guaranteed over shut down and restart operations. If the state is not properly stored, a process known as checkpointing, restarting a device can lead to memory inconsistencies and incorrect operation \cite{cleancutcolin2018termination}. Additionally, the efficiency of these shut down and restart operations becomes critical, as they take away precious energy from operations that enable 
forward progress.
Also critical, it has to be ensured that forward progress can be made during phases of power-on time. If the energy required between two checkpoints is too large, the device will be unable to complete the computation. This results in a  program getting stuck repeating the same computation, which is referred to as non-termination. Thus, effective energy harvesting devices must have efficient techniques which enable correctness and forward progress, all while remaining within a modest hardware budget.

A recently proposed spintronic processing-in-memory (PIM) substrate, CRAM \cite{cram}, is uniquely well suited for energy harvesting applications. Operations on CRAM are highly energy efficient, enabling a low power budget. Further, as it is a PIM solution, it removes the need for energy hungry  data transfers between processor logic and (volatile) memories. 
{The main advantage, however, is that progress is automatically saved after every operation. CRAM consists entirely of non-volatile devices and the results of all computation are immediately stored in permanent memory. As there are very few variables required to maintain the architectural state, these can also be saved after each operation with minimal energy cost. Effectively, checkpointing occurs after every operation. 

Checkpointing after each operation is not a new idea \cite{considereverycyclema2015architecture}, and for most systems this would generally be considered inefficient \cite{cleancutcolin2018termination}. However, as CRAM is a non-volatile PIM substrate, most of the checkpointing operations come \emph{for free}. Hence, CRAM can restart a program from the very last operation with fast and efficient shut down and restart. Additionally, CRAM is always in a state that can be recovered from. The power can be cut \emph{instantly and unexpectedly}, and it will still restart correctly. The maximum penalty is repeating the last instruction. We refer to this capability as \emph{instantly restart-able.} This provides a significant advantage, as shut down and restart procedures for more conventional energy harvesting devices introduce additional latency and energy, and significant complexity. 

While PIM has been used previously in energy harvesting devices \cite{RRAMsu2017462gops}, in such cases the PIM array acts as a sub-component of the system, leaving much of the computation to an external processor. Hence, these systems do not exploit the full potential of PIM as CRAM does. 
Other non-volatile PIM substrates such as \cite{pinatubo}, which could potentially be adapted similarly, use external logic at the periphery of the memory array (including sense amplifiers) for computation, which is not only less energy efficient, but also  makes adaptation for intermittent processing more complex.}

In this paper, we introduce \name\ (\textbf{M}achine Learning \textbf{A}ccelerator in \textbf{ST}T-MRAM for \textbf{E}nergy Ha\textbf{R}vesting Applications) which is built using CRAM \cite{cram}. While based on CRAM, \name\ has a different cell design which reduces energy consumption during computation. As a case study,
we implement support vector machines (SVM), which are widely used machine learning algorithms. We demonstrate how \name\ can provide high performance and energy efficiency on such applications while also having efficient shut down and restart procedures. Additionally, we show how another modification to the CRAM cell, the addition of a spin-hall effect (SHE) channel, can further increase energy efficiency. In Section \ref{sec:background} we 
provide 
the working principles of CRAM. In Section \ref{sec:SVM} we describe the the support vector machine we use as an application. We introduce design specifics of \name\
in Section \ref{sec:accelerator} and show how we guarantee correctness in Section \ref{sec:correctness}. We set up the evaluation in Section \ref{sec:evaluation}, show our results in Section \ref{sec:results}, discuss related work in Section \ref{sec:relatedwork}, and conclude in Section \ref{sec:conclusion}.

\section{Processing In Spintronic Memory}
\label{sec:background}
Spintronic memory in the form of STT-MRAM is an emerging technology, with a few products already commercially available \cite{everspin}. Due to its non-volatility, high density, speed, and endurance, STT-MRAM is being considered as a universal memory replacement \cite{dong2008circuit}. STT-MRAM arrays use one magnetic tunnel junction (MTJ) and one access transistor per cell. \name\ maintains the same basic cell structure. By making light modifications to the array, we are able to connect MTJs in such a way to enable logic operations to be implemented within the array. Therefore, \name\ is capable of being used as both a standard STT-MRAM array and as a computational substrate. \name\ is unique in that the computation does not require any external logic circuits or the use of sense amplifiers, making the computation contained \emph{entirely} within the array. In the following, we explain MTJ basics and show how they can be used in logic operations. Then we demonstrate how these operations can be performed within the array structure.

\subsection{Magnetic Tunnel Junction (MTJ)}
{STT-MRAM arrays are built with magnetic tunnel junctions (MTJ). The MTJ is a resistive memory device which consists of two magnetic layers (fixed layer and free layer) which are separated by an insulator. The polarity of the free layer can change but the fixed cannot. When the fixed and free layers are aligned, the MTJ is in the parallel (P) state, which has a low resistance and corresponds to logic value 0. When the layers are opposing, the MTJ is in the anti-parallel (AP) state, which has a high resistance and corresponds to logic value 1.

The state can be determined by applying a voltage across the device and sensing the amount of current that travels through it. If a sufficient amount of current is driven through the device, it will change state. Importantly, the state it changes to depends on the direction of the current. This is key to our ability to ensure correctness in spite of power outages. When current flows from the free layer (fixed layer) to the fixed layer (free layer), it switches the MTJ to the AP (P) state. }

\subsection{Logic Gates}
\label{sec:gates}
\begin{figure}
    \centering
    \includegraphics[scale=0.23]{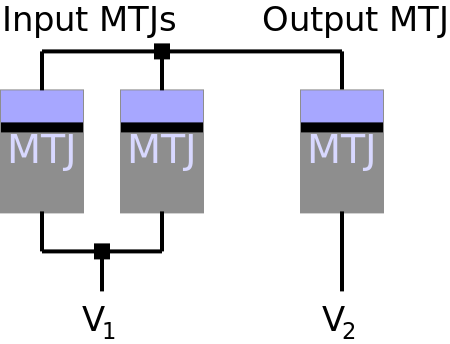}
    \caption{MTJs connected to implement a 2-input logic gate. The preset value of the output MTJ and the polarity and magnitude of the voltage applied between $V_1$ and $V_2$ determines the type of logic gate. The fixed layer is colored in grey and the free layer in light blue.}
    \label{fig:logic}
    \vspace{-.1cm}
\end{figure}

Before showing how logic can be implemented in the \name\ array, we demonstrate how logic gates are performed on MTJs. The configuration for a two-input logic gate is shown in Figure \ref{fig:logic}. The two MTJs in parallel are the inputs to the logic gate, and the MTJ in series with them is the output. The output must be preset to a known value. For example, the output is preset to 0 (low resistance) for a NAND gate. To implement a NAND gate, a voltage is applied across the two terminals, $V_1$ and $V_2$, such that current flows from the input MTJs to the output MTJ. If either of the input MTJs is 0 (low resistance) there will be sufficient current to switch the output MTJ to 1. If both input MTJs are 1, there will be insufficient current to change the state of the output MTJ, and it will remain at 0. Therefore, the state of the output MTJ follows the truth table for a NAND gate, it is 0 only if both inputs are 1.{ 
Due to the underlying physics, MTJ switching depends on the direction of the current. Current flowing from the input MTJs to the output MTJ can only cause the MTJ to switch to 1. It cannot cause it to switch to 0. }

{All other logic gates are performed similarly. In order to implement other gates, we can change the number of inputs, the preset value of the output, or the direction of the current. For example, using the same circuit (same number of inputs), we can perform an AND gate on the two inputs. In this case, the output MTJ is preset to 1 and current is applied in the opposite direction. This is because we want the output MTJ to switch from 1 to 0 (rather than 0 to 1) if either of the input MTJs is 0 (low resistance). Hence, current flows from the output MTJ to the input MTJs, and if either of the inputs MTJs is 0, there is sufficient current to switch the output to 0. This follows the logic of an AND, where the output will be 0 if either of the inputs is 0. }


Many other common gates can be implemented in this way, such as NOT, COPY, and (N)OR. (Inverse) Majority gates with more inputs are also possible.
However, the difference in the combined resistances of different inputs gets harder to distinguish for more inputs as more resistances are added in parallel. Generally, two input gates are robust as MTJ resistance is much larger than the parasitic resistance of the access transistors. Hence, resistance differences between different input combinations are also large. We restrict ourselves to a maximum of two inputs per gate in our evaluation.

More complex operations are broken down into these basic logic operations. For example, a full-add can be performed with 9 NAND gates and 7 temporary bits. To perform a full-add in \name, we perform the 9 NAND gates sequentially and use spare MTJs to hold the temporary bits. {Using full-adds, full-subtracts, and other primitive operations we can perform integer or fixed-point arithmetic, thus enabling us to implement our benchmarks. Naturally, the latency for each complex operation is quite high, as it must be broken down into its constituent gates which are then performed sequentially. However, as we will show in later sections, this can be compensated for by performing many data independent operations in parallel. }

\subsection{Array Architecture}
\begin{figure}
    \centering
    \includegraphics[scale=0.15]{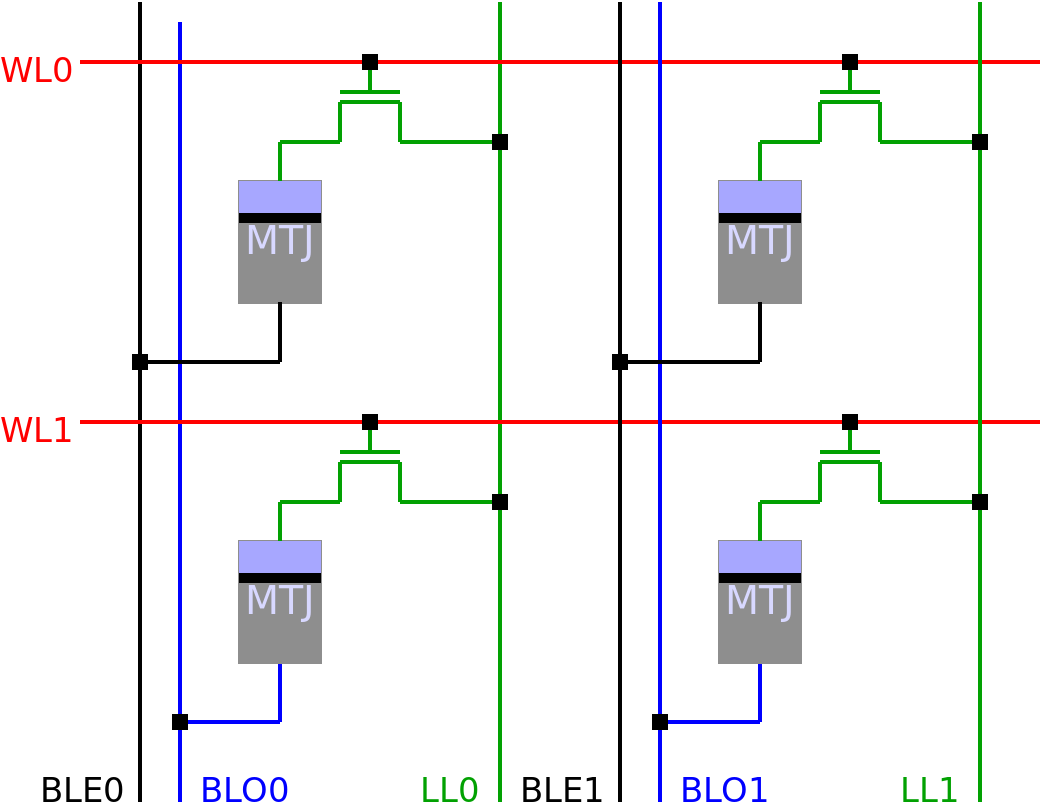}
    \caption{Four cells in two columns and two rows of 1TM configuration. Abbreviations are Wordline (WL), Logic Line (LL), Bitline Even (BLE), Bitline Odd (BLO). LL enables the MTJs to be connected, with voltages applied over BLE and BLO.}
    \label{fig:1T}
\end{figure}

\name\ {is an STT-MRAM array with some additional hardware. Four cells located in adjacent rows and columns are shown in Figure \ref{fig:1T}. Each memory cell consists of one MTJ and one access transistor. In each column there are two bit lines, bit line even (BLE) and bit line odd (BLO), and a \textit{logic line} (LL). In each row there is a wordline (WL) that controls the access transistor. Each MTJ is connected to the LL through the access transistor and to one of the two bit lines. Cells in even rows are connected to BLE and cells in odd rows are connected to BLO. We now describe how memory and logic operations are performed in the array.}
 


\noindent \textbf{Memory Operation:} To read or write from row $n$, activate WL$n$ and apply a voltage differential across LL and the bitlines. Current will only travel through the bitline with the same parity as $n$. Current can be sensed on the bit lines to perform a read, or a large current can be driven through the MTJ to perform a write.

\noindent \textbf{Logic Operation:} To perform a logic operation with inputs on rows $n_1$ and $n_2$ with output in row $m$, preset row $m$ by performing a write operation. Activate WL$n_1$, WL$n_2$ and WL$m$. Apply a voltage differential across BLE and BLO. Current travels from one bit line, through the MTJs in rows $n_1$ and $n_2$, through the LL, through the MTJ in row $m$, and back to the other bit line. Depending on the states of the MTJs in rows $n_1$ and $n_2$, the state of the MTJ in row $m$ will either change state or not. $n_1$ and $n_2$ must have the same parity and $m$ the opposite. 

{Voltage which drives the operation is applied to every column in which the specified operation should take place. The peripheral circuitry determines which columns these are, which can be specified by dedicated instructions as will be described in Section \ref{sec:instructions}. Hence, while only one operation can be performed in a column at a time, an operation can be performed in many columns simultaneously. This gives \name\ \emph{column level parallelism}. {This bears some resemblance to bit-serial architectures.}}

\subsection{Spin Hall Effect Channel}
\label{sec:she}
\begin{figure}
    \centering
    \includegraphics[scale=0.15]{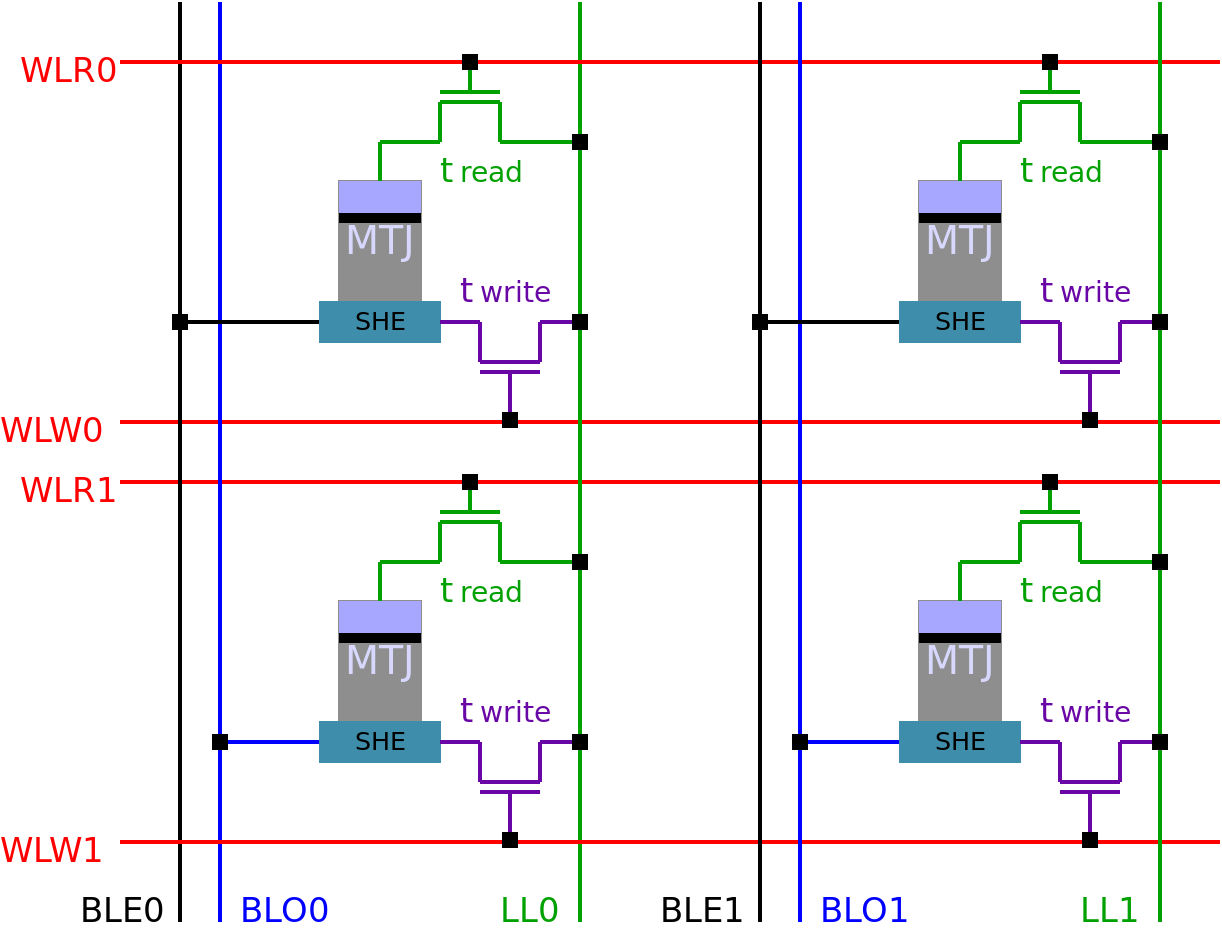}
    \caption{Four cells in two columns and two rows of 2T-1M SHE \name\ configuration. There are two wordlines, word line read (WLR) and word line write (WLW).}
    \label{fig:she}
\end{figure}

Augmenting each \name\ cell with a spin hall effect (SHE) channel can further improve energy efficiency. Four augmented cells in two rows and two columns are shown in Figure \ref{fig:she}. Each device (MTJ and combined SHE channel) now has three terminals, instead of two. This necessitates the addition of a second transistor per cell. One end of the SHE channel is connected directly to the bit line. There are two word lines per row, word line read (WLR) and word line write (WLW). Each controls one of the access transistors. The access transistor $t_{read}$ is controlled by WLR and connects the other end of the MTJ to the logic line. When $t_{read}$ is activated, current passes through the SHE channel and the MTJ device. This allows the MTJ state to affect the current that travels through it. This is used when reading the MTJ state and when the MTJ is used as an input to a logic operation. $t_{write}$ is controlled by WLW and connects the other end of the SHE channel to the logic line. When $t_{write}$ is activated, current only passes through the SHE channel. This current, while not effected by the state of the MTJ, can still change the state of the MTJ. This configuration is used when writing to the MTJ or when the MTJ is the target output of a logic operation.

The SHE channel has a few benefits. One is that the required current density to induce switching in the SHE channel is lower, allowing for a reduction in the energy of write and logic operations. It also removes the need to preset the value of the output MTJ for logic operations, as the state of the MTJ does not affect the SHE channel resistance, and hence does not need to be accounted for. This saves latency and energy by making many write operations unnecessary during the run of a program.

\section{Support Vector Machines}
\label{sec:SVM}
To show the capability of \name, we implement Support Vector Machines (SVM) as a case study. SVMs are widely used machine learning algorithms. Currently, they are second in popularity to neural networks. The applications for neural networks and SVMs overlap considerably, however they offer different advantages. Neural networks generally provide higher accuracy at a cost of more complexity. They are particularly good at image recognition, which has resulted in increased attention in the last few years. 

{\name\ can also serve as an accelerator for neural networks, as it is capable of performing a universal set of logic operations, hence, any program. Compressed neural networks \cite{InferenceBeyondEdge} and binary neural networks \cite{bnn} would also be well suited for the energy harvesting domain. SVMs are effective and simple classifiers for typically smaller data sets, which we chose as a case study in this paper, without loss of generality. Particularly, we found SVMs to perform well on image recognition and human activity recognition. }However, there is a trade-off, as SVMs can struggle with some problems. For example, we were unable to achieve reasonable accuracy on the speech recognition data set, which neural networks have performed well on \cite{InferenceBeyondEdge}. {Generally speaking, whether SVMs or neural networks are a superior choice depends on the target problem, but both are applicable for energy harvesting applications.}


SVMs work by mapping inputs to a higher dimensional space, where the different classes become linearly separable from each other. Training an SVM involves finding a set of training inputs (support vectors) and weights (coefficients) which are good indicators of a particular class output. New inputs are then compared to the chosen training inputs, and whichever class it is most similar to is the assigned class. 

For all benchmarks we use a polynomial kernel with a degree of 2. For inference, the main computation is effectively computing the dot product between an input vector and each of the support vectors. The results of these dot products are then squared, multiplied by the coefficients, and finally added together. By design, SVMs have two class outputs, where the sign of the output value is the classification. 

In this work, we opt for the simplest extension to multi-class problems: we train a separate SVM for each possible output class. Each SVM has the task of identifying its assigned class. For example, MNIST has 10 different classes for digits 0-9. We train 10 SVMs each identifying each digit. The output is 10 scores for ``how similar'' the input is to each digit. We take the highest output of the 10 classifiers to be the final classification. Our SVMs are custom designed, however we compare our results with libSVM \cite{libsvm} and achieve comparable accuracy. We perform training offline in software and only consider inference acceleration on \name.


\section{\name\ Accelerator}
\label{sec:accelerator}
Energy harvesting systems are powered by their environment. If the environment does not provide enough power, the system will have to accumulate energy over time and consume it in bursts \cite{InferenceBeyondEdge}. Therefore, such devices must consume as little energy as possible and be capable of tolerating power outages while maintaining program correctness. \name\ is a natural fit for such a paradigm as logic operations are highly energy efficient and the memory is entirely non-volatile. Additionally,  all computation occurs within the memory so progress is effectively saved after each operation. This greatly simplifies strategies to maintain correctness. In this section, we detail a basic \name\ design which is tightly tailored to energy harvesting applications.

\subsection{Hardware Organization}
\begin{figure}
    \centering
    \includegraphics[scale=0.37]{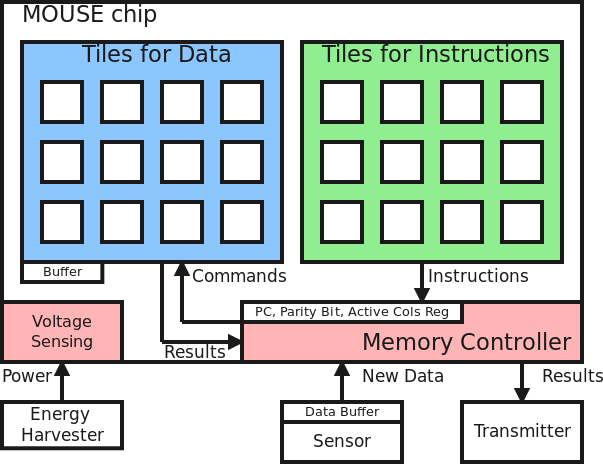}
    \caption{Overview of \name. 
    \name\ tiles hold data and instructions. The memory controller fetches instructions and broadcasts them to the tiles. The memory controller is also responsible for maintaining the program counter and valid bits to preserve architectural state. }
    \label{fig:network}
\end{figure}

\name\ has a tiled architecture. Certain \name\ tiles are dedicated for instructions, while all others are dedicated for data and computation, as shown in Figure \ref{fig:network}. \name\ has a larger storage capacity than is typical for energy harvesting devices. This is due to two reasons. First, STT-MRAM is dense and has extremely low standby power, giving the memory a low area and energy impact. {For example, NVSIM \cite{nvsimdong2012nvsim} reports the size of 64MB STT-MRAM array, which is nearly twice the size of our largest configuration, as 15.12 \SI{}{ \milli \meter }$^2$. A 256-MB STT-MRAM memory device manufactured by Everspin \cite{everspin} comes in a package that is 130 \SI{}{ \milli \meter} $^2$. For reference, the MSP430FR5994 micro-controller commonly used as a sub-component of energy harvesting systems \cite{InferenceBeyondEdge,mcu1colin2018reconfigurable,mcu2hester2016amulet,mcu3hester2015tragedy,mcu4hester2017flicker,mcu5sample2008design} is over 100 \SI{}{ \milli \meter}$^2$.} Second, as there is no need for external processor logic or area costly volatile memory (such as SRAM), and due to minimal peripheral circuitry, nearly the entire area budget is available for memory arrays.  There are only five components of \name\ that are not memory arrays:
\begin{enumerate}
    \item A memory controller that reads instructions from the instruction arrays and issues all instructions; 
    \item An 128B memory buffer that facilitates communication between \name\ tiles;
    \item A non-volatile register for program counter;
    \item A non-volatile register for storing a single instruction;
    \item Voltage sensing circuitry for monitoring the power source.
\end{enumerate}

The memory controller only needs to differentiate between three instruction types as will be described in Section \ref{sec:instructions}. All computation and memory operations are performed in the tiles, hence the controller needs only broadcast the appropriate command to the tiles. The memory buffer is the same size as one line of the \name\ tiles and is used for intermediate storage when transferring data between tiles. The non-volatile registers are used for maintaining correctness during power outages, as will be described in Section \ref{sec:intermittent}. The voltage sensing circuitry is standard for energy harvesting systems, and is as described in \cite{processor}. 

\subsection{Instructions}
\label{sec:instructions}

\begin{figure}
    \centering
    \includegraphics[scale=0.25]{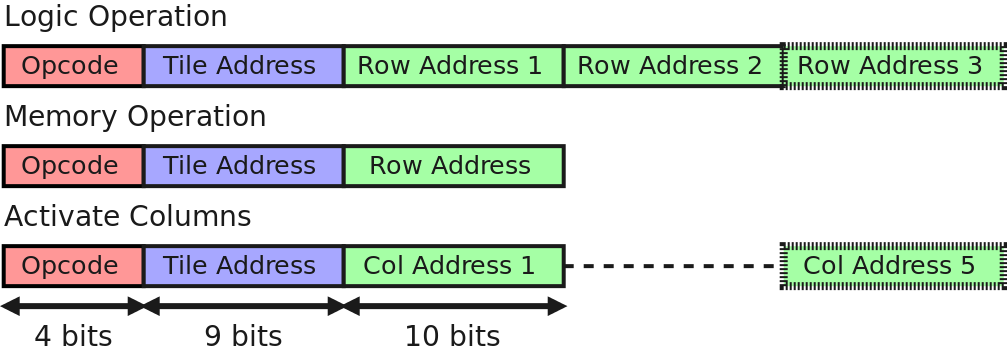}
    \caption{\name\ instruction formats. There are three types of instructions, logic, memory, and an additional activate columns instruction for configuration. Opcodes are 4 bits; tile addresses, 9 bits; and row and column addresses, 10 bits each. Dashed items are optional.}
    \label{fig:instructions}
\end{figure}

Instructions for \name\ are 64-bit and the formats are shown in Figure \ref{fig:instructions}. There are three types of instructions, logic operations, memory operations, and column activation. Memory operations are the same as standard read and write operations for STT-MRAM. Instructions for logic operations specify the type of operation (which determines the applied voltage level) and the rows on which input and output cells reside. When a logic instruction is issued, it will be applied to every column that is currently active. Columns are activated by the \textit{Activate Columns} instruction, which provides a list of column addresses {to a column decoder}. Once columns are activated they are held active by a latching mechanism as proposed by \cite{pinatubo}. This allows columns to remain active over multiple instructions. As columns need to be changed infrequently, typically staying active for many instructions, the peripheral cost for activation is amortized. This cost is further reduced by modifying the encoding to allow for bulk addressing, similar to the procedure in \cite{ambit}. 

Compiling instructions for \name\ is non-trivial as it requires some knowledge of the hardware to make efficient use of potential parallelism. This situation is analogous to compiling for GPU architectures from Open-CL or CUDA code. Unfortunately there is no equivalent for PIM. In our work the instructions are custom generated, however, the architecture and data layout for \name\ is similar to a number of other processing-in-memory (PIM) substrates \cite{pinatubo,ambit}. 

{Some tiles are dedicated to store the instructions. The instructions are written into these tiles before deployment. Once active, the memory controller fetches each instruction from the instruction tiles, decodes it, and then broadcasts it to the tiles storing data. Instructions vary in the amount of time they take to complete. This is because specifying row and column addresses has an associated latency, and different instructions have different numbers of addresses. Logic operations can use 2 or 3 rows and column activation can specify up to 5 columns. To ensure that every instruction finishes, the memory controller waits longer than the longest instruction before issuing the next. This does not impact performance as, due to power restrictions of energy harvesting sources described in Section \ref{sec:powerdraw}, \name\ does not issue instructions as fast as possible. Hence, this wait period can use already existing spare time.  }

{In this work, as we are only performing inference, the instructions performed are not input dependent. 
Instructions are performed in sequential order until the program repeats. 
We provide more detail on issuing instructions in Section \ref{sec:correctness}}.

\subsection{Power Draw}
\label{sec:powerdraw}
While \name\ operations are very energy efficient, \name\ can still consume a lot of power due to large amounts of parallelism. If unconstrained, \name\ can consume up to approximately 15 \SI{}{  \milli \watt}. Unfortunately, typical energy harvesters can only provide up to a few hundred micro watts of power \cite{processor}. Fortunately, \name\ can be easily configured to consume much less power at the cost of performance. The trick is to reduce parallelism and perform more operations sequentially. However, we choose to reduce the rate at which we issue instructions so operations are performed at a lower frequency. This introduces idle time between instructions. This idle time can be used to perform other useful tasks, such as updating the architectural state.

\subsection{Intermittent Processing}
\label{sec:intermittent}
As energy harvesting systems frequently experience power outages, they must be designed to perform intermittent processing. This involves addressing the challenge of maintaining correct state while repeatedly shutting down and restarting. The mechanism for maintaining state also need be efficient, as to avoid consuming the precious energy available for program execution. A number of techniques have been designed to ensure correctness \cite{cleancutcolin2018termination,coatiruppel2019transactional,chinchillamaeng2018adaptive,dnngobieski2018intermittent}. These studies have devised sophisticated techniques to ensure correctness while introducing minimal backup and restart overhead. In contrast, \name\  maintains correctness with just a program counter (PC){ and an additional non-volatile status bit.} While extremely simple, and would be crude for other architectures, it is a natural fit for \name. More sophisticated techniques are unsuitable and unnecessary as \name\ has no volatile data to backup. As \name\ performs all computation within the non-volatile memory, progress is saved after each operation. This makes restarting after the last instruction possible and ideal. 

When \name\ restarts, only two pieces of information are required: the last instruction that was performed and the columns that were active. In order to restart from the last instruction, \name\ writes the PC into a non-volatile register after each instruction. When \name\ gains sufficient power to restart, it simply reads the next instruction from the address in the PC. 
In the worst case, the power is cut after the last instruction is issued and performed, but before the update to the PC register. {This does not break correctness as the same result is obtained if a single instruction is repeated multiple times, meaning it is \emph{idempotent}, as will be shown in Section \ref{sec:correctnessgate}. The only requirement is that the PC update happens strictly after each instruction is performed.  Restarting after the very last instruction not only minimizes the amount of work potentially lost on shutdown, but it simplifies the restart process. The simple correctness guarantee, an operation being \emph{idempotent}, does not hold if we were to repeat multiple instructions. This is because over the course of multiple instructions, multiple temporary values can be created. These temporary values may be used later in the computation or periodically overwritten. Repeating multiple instructions on startup would require some method for ensuring correctness of these temporary values, such as performing additional presetting operations. This is certainly possible to do, but it introduces additional complexity. }

The second requirement is to restore the previously active columns, for which we use a similar procedure. Whenever an \textit{activate columns} instruction is issued, it is stored in  an additional instruction register. Reissuing this last \textit{activate columns } instruction is the first action on restart.
This scheme gives \name\ minimal backup and restart overhead. The cost is 1) continuous update of the program counter and \textit{activate columns} registers and 2) an additional issue of an \textit{activate columns} instruction on every restart. {Both of these actions incur far less energy than a typical logic instruction}. It is noteworthy that \name\ is always in a state which is safe to shut down in. Hence, \name\ maintains correctness even if power is cut unexpectedly.

We make sure that operations happen in the correct order by performing them sequentially; updates to (architectural) state maintaining registers occur only after the current instruction is performed. 
If run at full speed, \name\ consumes more power than a typical energy harvesting source can provide. This requires us to reduce the rate at which we issue instructions. Hence, there is already a time slack between instructions, during which these updates to the architectural state can be performed. 

\subsection{{System Integration}}
\name\ holds all static data required and performs all the computation. To be integrated into an energy harvesting system, \name\ needs to receive energy from an energy harvester, receive input from a sensor, and send output to a transmitter. In this work, we assume input data is stored in a non-volatile buffer in the sensor prior to inference. The sensor's buffer is assigned a tile address and is treated as one of the tiles. Additionally, the buffer contains a non-volatile valid bit indicating that new input is ready. When \name\ is ready for new input, the memory controller can check the valid bit and trigger a memory transfer. The memory transfer then consists of reads from the buffer and writes to the \name\ data tiles. These reads and writes are controlled by instructions at the beginning of the program. When \name\ finishes inference, the memory controller reads out the data from the tiles. This data is then available to be transferred to transmitter. In this work, we focus only on the accelerator and do not consider any overhead for the sensor or transmitter.

\section{Correctness Guarantee}
\label{sec:correctness}
We show that correctness is guaranteed in spite of power outages, even when unexpected. There are two components, the correctness of individual operations when interrupted or re-performed and correctness of state variables in transitions between states.

\subsection{Operation Level Correctness}
\label{sec:correctnessgate}
In this section we show that correctness is maintained if a single operation is repeated, meaning it is \emph{idempotent}. Given that the power may be cut at any moment, we must consider what happens when an operation is interrupted in all its possible stages. Since all operations in \name\ are threshold operations, the two stages are pre- and post-switching. Additionally, switching of the output MTJ either should or should not occur depending on the inputs. To be explicit, we use AND as an example, however, our observations here apply to all gates. 

The preset value for the output of an AND gate is 1, meaning the MTJ has a high resistance. { During operation, current is applied in a direction that could change the output state to 0. If either of the two inputs is 0, there will be a sufficient current to change the state, otherwise it will remain at 1.} We show the four possible cases in Table \ref{tab:correctness}. {If, due to the inputs, the output is not supposed to switch, the output MTJ will not switch before the power is cut or after the power is restored. On the other  hand, if the output is supposed to switch, it does not matter if it switches before the power outage or after. If the output MTJ does not switch before the power outage, it will switch once power is restored and the operation is re-performed. If the output MTJ does switch to 0 before the power outage, re-applying the power afterwards will leave the output at 0. This is because the direction of the current can only change the output to 0, it cannot revert it back to 1.} 

\begin{table}[]
    \centering
    \resizebox{0.5\linewidth}{!}{

    \begin{tabular}{c||p{3cm}|p{3cm}|}
         & Did not switch prior & Did switch prior  \\
         \hline
         \hline
         Should not switch & No values were changed. Repeating the operation is
         exactly the same as first time, no switching will occur. & Not possible. There was insufficient current to induce switching at all points during the operation, regardless of interruption
 induced switching. \\
         \hline
         Should switch & No values were changed. Repeating the operation is as performing for first time, and will now result in correct operation.  & The output has already switched to 0. Reapplying voltage will result in a larger current, however the direction of the current can only result in switching the output to 0. Hence, this is analogous to applying the voltage for a gate for a longer duration. \\
    \end{tabular}
    }
    \caption{Four possible cases for re-performing an interrupted AND gate. The output MTJ either should or should not switch for correct operation, and it either did or did not prior to the power being cut.}
    \label{tab:correctness}
\end{table}

Putting it  all together, the basic idea is that repeating a logic gate is effectively the same as performing the gate for a longer duration. Doing so results in an identical outcome, regardless of whether the output MTJ switched before interruption or not. The case for writes is simpler. The result of a write operation does not depend on the preset value, hence repeating a write is effectively writing the value twice. Power interruptions can lead to wasted energy, by re-performing unnecessary work, but do not result in corruption of logical values. 

\begin{figure*}
    \centering
    \includegraphics[scale=0.3]{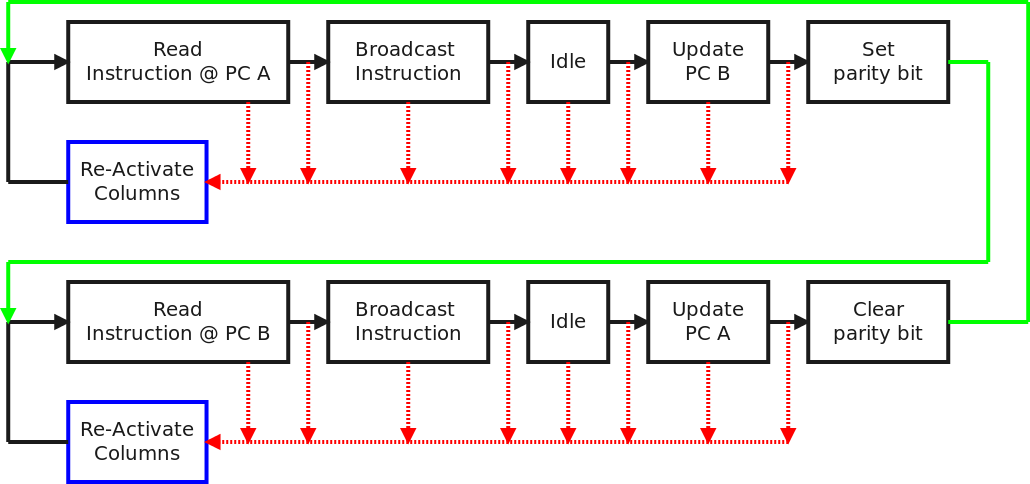}
    \caption{State transitions to maintain correctness. The program counter (PC) is duplicated and labelled A and B. Interrupts are highlighted in red, corrective measures in blue, and forward progress in green. Individual instructions are safe to re-perform, as detailed in Section \ref{sec:correctnessgate}}
    \label{fig:flow}
\end{figure*}

\subsection{Maintaining Correct State}
\label{sec:correctnessstate}
It must also be ensured that the memory controller can tolerate unexpected interruptions. The memory controller reads instructions from the address held in the non-volatile program counter (PC), decodes them, and broadcasts them to the data tiles. It then updates the PC.  If power is cut during a write operation to the PC, the value may be corrupt. We solve this by duplicating the PC register and maintaining a parity bit. If the parity bit is 0 then PC-A is valid and if the parity bit is 1 then PC-B is valid. The valid PC register points to the instruction currently being executed. After an instruction is completed, the value stored in the valid PC register is read, updated, and written to the invalid PC register. The new PC value now points to the next instruction that is to be executed. After the PC register update, the parity bit is flipped. This process is depicted in Figure \ref{fig:flow}. With this scheme, a valid copy of the PC is maintained at all times. {If power is cut after the update to the invalid PC but before the parity bit is flipped, the memory controller will consider the old PC to be valid on restart. This results in the previous instruction being re-performed. This does not introduce errors as individual instructions are idempotent, as shown in Section \ref{sec:correctnessgate}. Hence, power can be cut at any point during the execution of an instruction and the memory controller can restart correctly.}

\section{Evaluation Setup}
\label{sec:evaluation}

\noindent {\bf Benchmarks:} Energy harvesting systems are ideal for applications in which the system is difficult or inconvenient to power directly or with batteries. Examples include remote sensors and wearable tech. We choose benchmarks which are representative of different possible use cases, along with an additional standard benchmark.

MNIST \cite{MNISTlecun1998gradient}, as an example image recognition for sensor networks, is a digit recognition data set, where there are 10 classes for digits 0-9. The input is a grey scale $28\times28$ pixel image with 8-bit precision. The pixels are placed row wise into a 784 element vector. We also use a binarized version, where pixels that are greater than a threshold value are set to 1 and others to 0. This allows us to replace multiplications with AND gates for some parts of the computation.

Human Activity Recognition (HAR) \cite{HARrequest}, as an example for wearable tech, is a data set containing measurements from an accelerometer and gyroscope embedded in a smartphone, which is carried by participants performing a variety of activities. The task is to classify each set of readings to which activity is being performed. We represent the input with fixed point integer format with 8-bit precision. Each input is a vector of 561 elements.

ADULT \cite{ADULTkohavi1996scaling} is a commonly used benchmark for SVMs that contains census information and the task is to classify whether an individual makes greater than \$50,000 per year or not. We use a reformatted version of the data set from libSVM \cite{libsvm}. Each input is a 15 element vector where each element is an 8-bit integer.


Our SVMs are trained and tested in R \cite{R}. They are custom designed, however we do compare our results with libSVM \cite{libsvm} with the same inputs and obtain similar accuracy. In our custom implementation we do not use any operations that would be inefficient in \name; all programs consist of bit-wise and integer arithmetic. 

\begin{table}[]
    \centering
      \resizebox{0.4\linewidth}{!}{
    \begin{tabular}{|c|c|c|}
         Parameter & Modern & Future \\
         P State Resistance & 3.15 \SI{}{ \kilo \ohm} & 7.34 \SI{}{ \kilo \ohm} \\
         AP State Resistance & 7.34 \SI{}{ \kilo \ohm} & 76.39 \SI{}{ \kilo \ohm} \\
         Switching Time & 3 \SI{}{ \nano \second} \cite{mtjsaida2016sub} & 1 ns \\
         Switching Current & 40 \SI{}{ \micro \ampere} \cite{mtjsaida2016sub} & 3 \SI{}{ \micro \ampere} \\
    \end{tabular}
    }
    \caption{Parameters for MTJ devices}
    \label{tab:mtj}
\end{table}

\noindent {{\bf Performance and Energy Model:}} We simulate the benchmarks on \name\ with an in-house simulator, also implemented in R. \name\ has a tiled architecture. We set each tile to have a capacity of 128KB, which is an 1024x1024 array. {We chose this size as it is a commonly recommended subarray size for non-volatile memories from NVSIM} \cite{nvsimdong2012nvsim}. We  experiment with both modern MTJs \cite{modernMTJsaida2016sub} and estimates of future MTJs \cite{she}. {Expected improvements in MTJ devices will drastically increase energy efficiency.} For future MTJs, we test both STT and SHE based architectures. The MTJ parameters we use are shown in Table \ref{tab:mtj}. For future MTJs, two techniques enable a reduction in the switching current, 1) decreasing the damping constant of ferromagnetic materials \cite{m1sato2014properties,m1mizukami2009low,m1durrenfeld2015tunable} and 2) using a dual-reference layer structure \cite{m2hu2015stt,m2diao2007spin}. To be conservative, we assume 3 \SI{}{ \micro \ampere }, however, switching currents as low as 1 \SI{}{ \micro \ampere\ }are possible. To estimate latency and energy cost due to peripheral circuitry, we take data from NVSIM \cite{nvsimdong2012nvsim} which reports results for modern STT-MRAM memories. We set our peripheral circuitry costs so that they consume the same percentage share of the total latency and energy as reported by NVSIM. In addition to the latency and energy required for performing the instructions, we also account for the overhead involved in reading the instructions from the arrays, updating the program counter and valid bits, storing the most recent \textit{activate columns} instruction, and the re-issuing of the last \textit{activate columns} instruction whenever the system restarts. 

We first evaluate the performance of \name\ under continuous power. We do not limit the power it consumes to allow it to achieve its maximal throughput. Then, we evaluate \name\ under energy harvesting conditions. Following the approach in \cite{processor}, we model the power source as a 16 kHz square wave with a duty cycle. The duty cycle is the percentage of the time the power source is on, e.g., a duty cycle of 0.5 means power is on half the time. We report results for each benchmark over a variety of duty cycles. Additionally, during power on time we need to keep the power within a realistic budget for energy harvesting systems, approximately a couple hundred micro watts \cite{processor}. To reduce power consumption we idle between instructions. This increases latency but ensures the power budget is within energy harvesting limitations. Following metrics provided in \cite{ehmodelsan2018eh}, we report energy dedicated to different components. In addition to total energy, we report Backup energy, Dead energy, and Restore energy. Backup is operations performed prior to shut down to save state. For us, this is the continual writing of the PC, parity bit, and storing each activate columns instruction in an additional instruction register. Dead energy is energy spent re-performing work that was lost during shut down, which in this case is repeating the last instruction on restart. Restore energy includes any operation needed to prepare \name\ for computation on restart. For us, this is issuing the most recent activate columns instruction.

\noindent {{\bf Area Overhead:}}
\name\ tiles have a similar area overhead as STT-MRAM arrays. \name\ has an extra bit line per column for the STT configuration. For the SHE configuration, it has an extra transistor and SHE channel for each cell. The impact of the additional bit line is minor but the additional transistor has significant overhead. To get estimates for area overhead for modern STT-MRAM, we take results directly from NVSIM \cite{nvsimdong2012nvsim} using 22 \SI{}{ \nano \meter\ } node size. This does not take into account the additional circuitry in \name. NVSIM only allows for memory sizes that are powers of two, thus we choose the minimum size for which each benchmark fits. To estimate area overhead for \name\ with future MTJs, we create estimates for a cell size assuming access transistors with 1\SI{}{  \kilo \ohm\ } resistance. The access transistors dominate the area overhead. This is for two reasons: 1) the MTJs and SHE channel can be placed on a separate layer from the access transistors and 2) the access transistors are much larger. Technology scaling will help reduce the size of the \name\ tiles, but this is counteracted by the additional hardware required. To estimate peripheral circuitry, we take NVSIM results for area efficiency and adjust our estimates by the same ratio. We find that the \name\ arrays are slightly larger than modern STT-MRAM arrays, as shown in Table \ref{tab:area}.  As the SHE design has twice as many access transistors, the cell area is approximately twice as large. 

\begin{table}[]
    \centering
    \resizebox{.5\linewidth}{!}{

    \begin{tabular}{|c|c|c|c|}
    \hline
         Benchmark (Array Capacity) & Modern & Future STT & Future SHE  \\
         \hline
         MNIST (64MB) & 15.12 & 17.10 & 34.20 \\
         \hline
         MNIST Binarized (8MB) & 2.11 & 2.34 & 4.68 \\
         \hline
         HAR (16MB) & 4.02 & 4.47 & 8.93 \\
         \hline
         ADULT (1MB) & 0.41 & 0.46 & 0.91 \\
         \hline
    \end{tabular}
    }
    \caption{Area required for \name\ for different benchmarks and configurations. Units are in \SI{}{  \milli \meter }$^2$. Modern results come from NVSIM \cite{nvsimdong2012nvsim} and Future results
    come from our conservative cell area projections.}
    \label{tab:area}
\end{table}

\section{Evaluation}
\label{sec:results}

\begin{table*}[t]
    \centering
    \resizebox{0.75\linewidth}{!}{
    \begin{tabular}{|c|c|c|c|c|c|c|}
        \hline
         Benchmark & Latency (\SI{}{ \micro \second} ) & Energy (\SI{}{ \micro \joule} ) & \# SV & I/D Mem (MB) & Area (\SI{}{ \milli \meter }$^2$) & Accuracy\\
         \hline
         \multicolumn{1}{c}{\textbf{\name\ }}\\
         \hline 
         MNIST & 2,137 & 22.49 & 11,813 & 4.5 / 30.0 & 17.10 & 97.55 \\
         \hline
         MNIST (Binarized) & 66.63 & 1.10 & 12,214 & 1.25 / 6.0 & 2.34 & 97.37 \\
         \hline
         HAR (integer) \cite{HARrequest,HARweb}& 1,068 & 7.62 & 3,293 & 2.25 / 10.0 & 4.47 & 94.57 \\
         \hline
         ADULT & 116.50 & 0.12 & 1,909 & 0.25 / 0.5 & 0.46 & 76.12 \\  
         \hline
         \multicolumn{1}{c}{\textbf{libSVM} \cite{libsvm}}\\
         \hline
         MNIST & 7,830 & 234,900 & 8,652 & - & - & 98.05 \\
         \hline
         MNIST (Binarized) & 19,037 & 571,116 & 23,672 & - & - & 92.49 \\
         \hline
         HAR (integer) & 1,701 & 51,042 & 2,632 & - & - & 93.69 \\
         \hline
        ADULT & 379 & 11,370 & 15,792 & - & - & 78.62 \\
        \hline
        \multicolumn{1}{c}{\textbf{SONIC }\cite{InferenceBeyondEdge}}\\
        \hline
        MNIST & 2,740,000 & 27,000 & NA & 0.256 & $>$ 100 & 99 \\
        \hline
        HAR & 1,100,000 & 12,500 & NA & 0.256 & $>$ 100 & 88 \\
        \hline
    \end{tabular}
    }
    \caption{Unconstrained \name\ (using STT design and future MTJ devices) and related work under continuous power. Unconstrained means power consumption may be higher than what an energy harvesting power source can provide. libSVM is implemented on Intel Haswell E5-2680v3 processor, SONIC \cite{InferenceBeyondEdge} is implemented on MSP430FR5994 microcontroller. }
    \label{tab:performance}
\end{table*}

Results for \name\ unconstrained by power limitations are summarized in Table \ref{tab:performance}. This is assuming the power source is always on and that the \name\ accelerator can draw as much power as needed, typically a few \SI{}{ \milli \watt }. Also reported are results for the same benchmarks performed using libSVM on a CPU and an energy harvesting system SONIC \cite{InferenceBeyondEdge} under continuous power. libSVM is run on a supercomputing cluster using Intel Haswell 5-2680v3 processors. To be conservative, for libSVM we account only for the processor power consumption and assume it operates at its idle power. SONIC uses a TI-MSP430FR5994 microcontroller and is powered by a Powercast P2210B energy harvester. \name\ shows significant energy efficiency advantages, and improved latency over other implementations. \name\ does require more memory than SONIC, however, we believe this to be reasonable given that \name\ is implemented in high density STT-MRAM and does not need external processing logic or area costly volatile memory.

\name\ benefits greatly from binarizing the MNIST input. One bit inputs enable us to replace multiplications with AND gates, which significantly reduces the amount of computation required. This comes at a small cost in accuracy. The libSVM implementation struggles on the binarized MNIST inputs, and attempts to increase accuracy by adding many more support vectors. This increases the latency and energy of inference.

\begin{table*}
\centering
 \resizebox{0.75\linewidth}{!}{
\begin{tabular}{|c||c|c|c|c|c|c|}
\hline
Duty Cycle & Total T & Restore T & Total E & Backup E & Dead E & Restore E \\
\hline
\multicolumn{1}{c}{\textbf{Modern STT}}\\
\hline
1 & 3974130 & 0 & 1385.91 & 1.07676 & 0 & 0 \\
\hline
0.25 & 13296900 & 786.879 & 1964.52 & 1.42353 & 461.057 & 116.662 \\
\hline
0.01 & 40768800 & 2360.45 & 3120.57 & 2.09251 & 1383.11 & 349.986 \\
\hline
\multicolumn{1}{c}{\textbf{Future STT}}\\
\hline
1 & 90289.6 & 0 & 22.4758 & 0.0210049 & 0 & 0 \\
\hline
0.25 & 362694 & 8.3658 & 22.7693 & 0.0214355 & 0.216738 & 0.0662481 \\
\hline
0.01 & 7511610 & 185.424 & 28.3498 & 0.0251971 & 4.49416 & 1.36556 \\
\hline
\multicolumn{1}{c}{\textbf{Future SHE}}\\
\hline
1 & 75719.2 & 0 & 4.62454 & 0.000234049 & 0 & 0 \\
\hline
0.25 & 304104 & 7.62692 & 4.72564 & 0.000238884 & 0.0446134 & 0.0544654 \\
\hline
0.01 & 6299750 & 167.182 & 6.66675 & 0.000280234 & 0.91849 & 1.12166 \\
\hline
\end{tabular}
}
\caption{Time (T) in \SI{}{ \micro \second} and Energy (E) in \SI{}{ \micro \joule} for MNIST  for different configurations and duty cycles.}
\label{tab:ehMNIST}
\end{table*}

\begin{table*}
\centering
 \resizebox{0.75\linewidth}{!}{
\begin{tabular}{|c||c|c|c|c|c|c|}
\hline
Duty Cycle & Total T & Restore T & Total E & Backup E & Dead E & Restore E \\
\hline
\multicolumn{1}{c}{\textbf{Modern STT}}\\
\hline
1 & 1047350 & 0 & 66.2254 & 0.29691 & 0 & 0 \\
\hline
0.25 & 3545990 & 394.563 & 93.4075 & 0.393292 & 21.5687 & 5.20246 \\
\hline
0.01 & 10871600 & 1183.91 & 148.104 & 0.571671 & 65.6813 & 15.6081 \\
\hline
\multicolumn{1}{c}{\textbf{Future STT}}\\
\hline
1 & 23882.5 & 0 & 1.09796 & 0.00579199 & 0 & 0 \\
\hline
0.25 & 97080.9 & 4.25462 & 1.11731 & 0.00598339 & 0.0103803 & 0.00296206 \\
\hline
0.01 & 2023320 & 94.3962 & 1.38619 & 0.00699701 & 0.220293 & 0.0609172 \\
\hline
\multicolumn{1}{c}{\textbf{Future SHE}}\\
\hline
1 & 20030.3 & 0 & 0.232652 & 6.50065e-05 & 0 & 0 \\
\hline
0.25 & 81429.9 & 3.82741 & 0.238355 & 6.71825e-05 & 0.00212102 & 0.00238624 \\
\hline
0.01 & 1698490 & 85.6235 & 0.328151 & 7.83354e-05 & 0.0452126 & 0.0490789 \\
\hline
\end{tabular}
}
\caption{Time (T) in \SI{}{ \micro \second} and Energy (E) in \SI{}{ \micro \joule} for Binarized MNIST  for different configurations and duty cycles.}
\label{tab:ehMNISTB}
\end{table*}

\begin{table*}
\centering
 \resizebox{0.75\linewidth}{!}{
\begin{tabular}{|c||c|c|c|c|c|c|}
\hline
Duty Cycle & Total T & Restore T & Total E & Backup E & Dead E & Restore E \\
\hline
\multicolumn{1}{c}{\textbf{Modern STT}}\\
\hline
1 & 1945340 & 0 & 469.4 & 0.532763 & 0 & 0 \\
\hline
0.25 & 6446870 & 428.981 & 664.945 & 0.695674 & 155.997 & 39.3572 \\
\hline
0.01 & 19766200 & 1287.27 & 1055.78 & 1.02001 & 467.79 & 118.072 \\
\hline
\multicolumn{1}{c}{\textbf{Future STT}}\\
\hline
1 & 44226 & 0 & 7.62075 & 0.0103929 & 0 & 0 \\
\hline
0.25 & 175988 & 4.59137 & 7.7174 & 0.0105 & 0.0736474 & 0.0223605 \\
\hline
0.01 & 3648560 & 102.033 & 9.60334 & 0.0123272 & 1.51944 & 0.46069 \\
\hline
\multicolumn{1}{c}{\textbf{Future SHE}}\\
\hline
1 & 37191.3 & 0 & 1.57293 & 0.000116298 & 0 & 0 \\
\hline
0.25 & 147952 & 4.13485 & 1.60668 & 0.000117508 & 0.015098 & 0.0183593 \\
\hline
0.01 & 3068790 & 92.595 & 2.26301 & 0.000137652 & 0.311281 & 0.378489 \\
\hline
\end{tabular}
}
\caption{Time (T) in \SI{}{ \micro \second} and Energy (E) in \SI{}{ \micro \joule} for Human Activity Recognition (HAR)  for different configurations and duty cycles.}
\label{tab:ehHAR}
\end{table*}

\begin{table*}
\centering
 \resizebox{0.75\linewidth}{!}{
\begin{tabular}{|c||c|c|c|c|c|c|}
\hline
Duty Cycle & Total T & Restore T & Total E & Backup E & Dead E & Restore E \\
\hline
\multicolumn{1}{c}{\textbf{Modern STT}}\\
\hline
1 & 189977 & 0 & 7.35651 & 0.0530281 & 0 & 0 \\
\hline
0.25 & 633798 & 72.4449 & 10.3876 & 0.0696842 & 2.40502 & 0.5831 \\
\hline
0.01 & 1943150 & 217.347 & 16.4139 & 0.101567 & 7.23325 & 1.74933 \\
\hline
\multicolumn{1}{c}{\textbf{Future STT}}\\
\hline
1 & 4330.44 & 0 & 0.120464 & 0.00103445 & 0 & 0 \\
\hline
0.25 & 17348.6 & 0.771005 & 0.122884 & 0.00105752 & 0.00155392 & 0.000330231 \\
\hline
0.01 & 361223 & 17.1016 & 0.151871 & 0.00123847 & 0.0238609 & 0.00682921 \\
\hline
\multicolumn{1}{c}{\textbf{Future SHE}}\\
\hline
1 & 3663.13 & 0 & 0.0263288 & 1.16688e-05 & 0 & 0 \\
\hline
0.25 & 14665.5 & 0.705909 & 0.0271286 & 1.19526e-05 & 0.000244041 & 0.000273212 \\
\hline
0.01 & 306041 & 15.4319 & 0.0373156 & 1.39622e-05 & 0.00506701 & 0.00563523 \\
\hline
\end{tabular}
}
\caption{Time (T) in \SI{}{ \micro \second} and Energy (E) in \SI{}{ \micro \joule} for ADULT benchmark for different configurations and duty cycles.}
\label{tab:ehADULT}
\end{table*}

We wish to specifically address the significant difference in performance between \name\ and SONIC \cite{InferenceBeyondEdge}. SONIC is implemented on a conventional, low performance microprocessor. That design is highly economical, makes use of very scarce memory capacity, uses currently commercially available hardware, and has been proven experimentally. Additionally, the authors note that there is room for significant improvement in the efficiency. While we are reporting a significant latency and energy advantage, \name\ is not yet ready for fabrication. MTJ based logic has been experimentally demonstrated \cite{wang2016magnetic}, however a full-scale CRAM array has not yet. Integration into an energy harvesting system is still a few years away. 

Now we consider \name\ in a more realistic energy harvesting scenario. The power consumption of the unconstrained configuration (a few \SI{}{ \milli \watt }) is unrealistic for energy harvesting, where power budgets typically are a few hundred micro watts. Thus, we slow down the rate at which instructions are issued to remain within the power budget. Following the approach in \cite{processor}, we model the energy harvesting power source as a 16 \SI{}{ \kilo \hertz} square wave with various duty cycles. \name\ can only operate when the power source is on, and remains idle when the power is off. When the power is restored, there is the start up task of re-activating the active columns with the \textit{activate columns} instruction. Results are shown in Tables \ref{tab:ehMNISTB} -- \ref{tab:ehADULT}. {The reported total time includes power off time.} A duty cycle of 1 means the system is continuously powered. The time required for each benchmark increases with decreasing duty cycle. This is due to two reasons. Naturally, more time is spent powered off and \name\ cannot make forward progress. Also, with a lower duty cycle there are more interruptions during the run of the program and hence more restart operations. This is reflected in the increasing Restore time. However, as the restore process is fast, the Restore time remains a small fraction of the total time. For STT at a duty cycle of 0.01, the Restore time is only 185.42 \SI{}{ \micro \second} compared to the 7,511,610 \SI{}{ \micro \second} required for completion of the program, considering time while powered off. Because the SHE design does not require presetting of the output MTJ, it requires fewer operations than the STT design, and hence, finishes faster. As a result, the SHE design experiences fewer power interruptions during the run of the program. Thus, SHE has a lower Restore time than STT.

With decreasing duty cycle energy does not significantly increase. This is because no energy is spent while powered off and \name\ has a highly efficient restart process. Backup energy is the continual writing of architectural state variables, Restore energy is the peripheral cost of column re-activation on restart, and Dead energy is due to the possible re-execution of the previous instruction on restart. Typically, energy will increase with decreasing duty cycle as there are more interruptions during the run of the program, leading to more restart operations. The energy cost of restart depends on where in the program progress was interrupted. The more columns that were active at the time of interrupt, the higher the Restore energy cost will be. Due to the previously mentioned reduction in time to finish, and consequently the number of interruptions, SHE has a lower Restore energy than STT. For the STT configuration the Dead energy is typically much larger the Restore energy. For example, the Dead energy is 4.49 \SI{}{ \micro \joule} whereas the Restore energy is only 1.36 \SI{}{ \micro \joule\ } on the MNIST benchmark at a duty cycle of 0.01. For SHE, Dead and Restore energy are similar, the Restore energy is a comparable 1.12 \SI{}{ \micro \joule} but the Dead energy reduces to 0.918 \SI{}{  \micro \joule } for the same configuration. This is because most of the Dead energy goes towards logic operations, for which SHE has a higher efficiency. Backup energy is small relative to both Dead and Restore energy, as this corresponds to writing only a few bits on every cycle. For future STT, the Backup energy is only 0.025 \SI{}{ \micro \joule} for a duty cycle of 0.01. Backup energy can increase with decreasing duty cycle due to the repeating of backup operations on restart, which happens if the previous backup operation did not finish prior to shut down.

{Restore time, Dead energy, and Restore energy are all zero for the case of a continuously powered system. This is because there are no power outages and, hence, never a need to restart the system or re-perform any potentially unfinished instructions. }

{As modern MTJs are less efficient than predicted future MTJs, \name\ must idle for longer between instructions when using them. As a result, \name\ has a higher latency than SONIC \cite{InferenceBeyondEdge} on the MNIST benchmark and a comparable latency on the HAR benchmark. However, even with modern MTJs, \name\ still provides an energy efficiency advantage, with 1,385.91 \SI{}{  \micro \joule} for MNIST (relative to 27,000 \SI{}{ \micro \joule}) and 469.4 \SI{}{ \micro \joule} for HAR (relative to 12,500 \SI{}{ \micro \joule}).}

 We note that the ASIC accelerator in \cite{pudiannao} is the most relevant comparison to \name\ at full performance. However, there are no absolute values for latency, energy, or throughput reported in \cite{pudiannao}. All results are reported relative to a GPU baseline, for which absolute values are not reported. {The only comparison we can make is that they consume 596 mW of power, not counting external memory. \name\ consumes approximately 15 \SI{}{ \milli \watt} when unconstrained. We believe \cite{pudiannao} has a lower latency than \name, as ASIC designs typically have a latency advantage over PIM.}

\section{{Related Work}}
\label{sec:relatedwork}
Non-volatile processors \cite{considereverycyclema2015architecture,ma2017incidental,processor} are uniquely designed for intermittent computing by integrating  non-volatile memory near the compute units. {Unlike \name, these devices have a structure similar to traditional CPUs.} The authors of \cite{processor} propose a system using a THU1010N non-volatile processor for energy harvesting applications. They describe trade-offs in designing such a system and demonstrate its capability on a number of smaller benchmarks. A non-volatile processor is presented in \cite{RRAMsu2017462gops} which features PIM components. There is a controlling CPU that performs logic and control. A few RRAM arrays are used to accelerate computing in neural networks. {In this case, the PIM is a sub-component of the system, which also contains more traditional logic circuitry.} SONIC \cite{InferenceBeyondEdge} uses compressed neural networks to perform inference on a TI-MSP430FR5994 microcontroller. SONIC is powered by a Powercast P2210B energy harvester which collects energy from a 3W Powercaster transmitter. This design can perform MNIST image recognition, Human Activity Recognition, and speech identification with high accuracy. {This work is significant as it developed methods to ensure correctness for machine learning applications on conventional hardware for intermittent systems, and was proven experimentally. }
Capybara \cite{capybaracolin2018reconfigurable} is a dynamic power delivery system. In energy harvesting applications, tasks can be capacity-constrained (i.e., need to perform a large computation without being interrupted) or temporally-constrained (i.e., need to be run at a specific time). These constraints have conflicting needs. Capacity-constrained prefers a large energy buffer so it can complete a longer task. Temporally-constrained prefers a small buffer that recharges quickly. Capybara uses a re-configurable hardware energy storage mechanism and a software interface that allows the specification of energy needs for different tasks. This gives the system the ability to satisfy the requirements of both kinds of tasks. {While we do not focus on the power delivery system in this work, systems such as Capybara could be used to optimally supply \name\ with power. }Hibernus \cite{hibernusbalsamo2014hibernus},  on  the other hand,  is a system that reactively hibernates and wakes up. 

A number of techniques have been developed to enable intermittent computation on more traditional hardware. CleanCut \cite{cleancutcolin2018termination} works with LLVM to compile programs with checkpoints. There is a difficult balance when creating checkpoints. Too many, and it wastes valuable energy and time. Too few is worse, where the required energy between two checkpoints is larger than the energy that can be stored in the energy buffer. Thus, the program will get stuck, which is called non-termination. Finding such non-terminating conditions is difficult to do by hand. CleanCut uses a statistical energy model  to find potential non-terminating paths. Chinchilla \cite{chinchillamaeng2018adaptive} attempts to get the best of both worlds with adaptive checkpointing. When compiling, Chinchilla inserts many possible checkpoints. When running, it keeps a timer and only performs a checkpoint if the timer has expired. If the device fails to checkpoint before power outage, the timer is set to half the value. This occurs until it has found an appropriate amount of time to go before performing checkpointing. It also opportunistically increases the timer at specified intervals in an attempt to increase performance by reducing the number of checkpoints. Coati \cite{coatiruppel2019transactional} developed methods to ensure correctness for concurrent execution and interrupts for intermittent systems. The What's Next intermittent architecture \cite{whatsnextganesan2019s} uses approximation to improve performance. Rather than all-or-nothing approach, What's Next computes approximate results and continually improves the output. If an acceptable output is achieved it will skip to processing the next input. This enables the device to process more inputs as it does not waste time and energy achieving unnecessary accuracy. Alternatively, if there is sufficient energy available, it can continue refining the output. {These works have developed sophisticated techniques to enable more traditional computation substrates to achieve accuracy and performance on intermittent systems. With \name, we are able to significantly simplify our strategy as the substrate has a natural immunity to power interruptions. }

The EH model \cite{ehmodelsan2018eh} facilitates early design space exploration for energy harvesting architectures. It helps  finding a good balance to achieve minimal overhead for allowing maximal forward progress. As noted by the authors of \cite{ehmodelsan2018eh}, energy harvesting systems can generally be divided into two types, multi-backup, which perform many backups between power outages, and single back-up, which only save state once before a power outage. Multi-backup systems include Mementos, \cite{mementosransford2011mementos}, DINO \cite{dinolucia2015simpler}, Chain \cite{chaincolin2016chain}, Alpaca \cite{alpacamaeng2017alpaca}, Mayfly \cite{mayflyhester2017timely}, Ratchet \cite{ratchetvan2016intermittent}, and Clank \cite{clankhicks2017clank}. Single-backup systems include Hibernus \cite{hibernusbalsamo2016hibernus++}, QuickRecall \cite{jayakumar2014quickrecall}, and many  others \cite{aouda2014incremental,balsamo2016graceful,berthou2017peripheral,liu2016lightweight,lukosevicius2017using}. { According to this categorization, \name\ fits under a multi-backup system as we are constantly saving the architectural state.}

PIM has been studied for non-volatile memories with Pinatubo \cite{pinatubo}, for DRAM with Ambit \cite{ambit}, and for SRAM with Neural Cache \cite{neuralcacheeckert2018neural}. These technologies are meant to be integrated into the memory hierarchy of traditional CPUs and have not been considered for energy harvesting applications. Ambit and Neural Cache are not suitable for energy harvesting as they are volatile technologies. Pinatubo could be adapted and used similarly as CRAM in \name. However, Pinatubo uses logic external to the memory array for some operations. This adds complexity as these circuits would need to be protected against errors in intermittent computing. Additionally, Pinatubo uses sense amplifiers to perform computation, which is less energy efficient than the logic operations in CRAM. 

A number of RRAM PIM technologies also exist \cite{rram2016,rram2016b,rram2017,rram2018}. However, the RRAM array is used as an accelerator as a sub-component of the system. Hence, there is much additional circuitry and logic that occurs outside the memory. This significantly increases the difficulty to adapt to intermittent processing. Additionally, many RRAM accelerators rely heavily on ADC units, which have a significant area and energy overhead.

\section{Conclusion}
\label{sec:conclusion}
In this paper we presented \name, a machine learning accelerator in (non-volatile) memory for energy harvesting applications. The requirements for energy harvesting applications are extreme energy efficiency, efficient shut down and restart procedures, and correctness during intermittent execution. \name\ provides all of these by having highly energy efficient logic operations with simple and effective shut down and restart procedures. The non-volatility combined with processing in memory provides a natural progress saving mechanism which demands very little overhead. By simulation, we demonstrated that such a device would provide significant latency and energy efficiency advantages over state of the art approaches, and is is a promising candidate to bring machine learning to new domains.

\bibliographystyle{plain}
\bibliography{references}

\begin{thebibliography}{10}

\bibitem{everspin}
https://www.everspin.com/supportdocs/EMD3D256M08G1-150CBS1, 2019.
\newblock Accessed: 2019-08-10.

\bibitem{HARrequest}
Davide Anguita, Alessandro Ghio, Luca Oneto, Xavier Parra, and Jorge~Luis
  Reyes-Ortiz.
\newblock A public domain dataset for human activity recognition using
  smartphones.
\newblock In {\em Esann}, 2013.

\bibitem{aouda2014incremental}
Faycal~Ait Aouda, Kevin Marquet, and Guillaume Salagnac.
\newblock Incremental checkpointing of program state to nvram for
  transiently-powered systems.
\newblock In {\em 2014 9th International Symposium on Reconfigurable and
  Communication-Centric Systems-on-Chip (ReCoSoC)}, pages 1--4. IEEE, 2014.

\bibitem{balsamo2016graceful}
Domenico Balsamo, Anup Das, Alex~S Weddell, Davide Brunelli, Bashir~M
  Al-Hashimi, Geoff~V Merrett, and Luca Benini.
\newblock Graceful performance modulation for power-neutral transient computing
  systems.
\newblock {\em IEEE Transactions on Computer-Aided Design of Integrated
  Circuits and Systems}, 35(5):738--749, 2016.

\bibitem{hibernusbalsamo2016hibernus++}
Domenico Balsamo, Alex~S Weddell, Anup Das, Alberto~Rodriguez Arreola, Davide
  Brunelli, Bashir~M Al-Hashimi, Geoff~V Merrett, and Luca Benini.
\newblock Hibernus++: a self-calibrating and adaptive system for
  transiently-powered embedded devices.
\newblock {\em IEEE Transactions on Computer-Aided Design of Integrated
  Circuits and Systems}, 35(12):1968--1980, 2016.

\bibitem{hibernusbalsamo2014hibernus}
Domenico Balsamo, Alex~S Weddell, Geoff~V Merrett, Bashir~M Al-Hashimi, Davide
  Brunelli, and Luca Benini.
\newblock Hibernus: Sustaining computation during intermittent supply for
  energy-harvesting systems.
\newblock {\em IEEE Embedded Systems Letters}, 7(1):15--18, 2014.

\bibitem{berthou2017peripheral}
Gautier Berthou, Tristan Delizy, Kevin Marquet, Tanguy Risset, and Guillaume
  Salagnac.
\newblock Peripheral state persistence for transiently-powered systems.
\newblock In {\em 2017 Global Internet of Things Summit (GIoTS)}, pages 1--6.
  IEEE, 2017.

\bibitem{chargeusechandrakasan2008next}
Anantha~P Chandrakasan, Denis~C Daly, Joyce Kwong, and Yogesh~K Ramadass.
\newblock Next generation micro-power systems.
\newblock In {\em 2008 IEEE Symposium on VLSI Circuits}, pages 2--5. IEEE,
  2008.

\bibitem{libsvm}
Chih-Chung Chang and Chih-Jen Lin.
\newblock Libsvm: A library for support vector machines.
\newblock {\em ACM transactions on intelligent systems and technology (TIST)},
  2(3):27, 2011.

\bibitem{cram}
Zamshed Chowdhury, Jonathan~D Harms, S~Karen Khatamifard, Masoud Zabihi, Yang
  Lv, Andrew~P Lyle, Sachin~S Sapatnekar, Ulya~R Karpuzcu, and Jian-Ping Wang.
\newblock Efficient in-memory processing using spintronics.
\newblock {\em IEEE Computer Architecture Letters}, 17(1):42--46, 2017.

\bibitem{chaincolin2016chain}
Alexei Colin and Brandon Lucia.
\newblock Chain: tasks and channels for reliable intermittent programs.
\newblock In {\em ACM SIGPLAN Notices}, volume~51, pages 514--530. ACM, 2016.

\bibitem{cleancutcolin2018termination}
Alexei Colin and Brandon Lucia.
\newblock Termination checking and task decomposition for task-based
  intermittent programs.
\newblock In {\em Proceedings of the 27th International Conference on Compiler
  Construction}, pages 116--127. ACM, 2018.

\bibitem{mcu1colin2018reconfigurable}
Alexei Colin, Emily Ruppel, and Brandon Lucia.
\newblock A reconfigurable energy storage architecture for energy-harvesting
  devices.
\newblock In {\em ACM SIGPLAN Notices}, volume~53, pages 767--781. ACM, 2018.

\bibitem{capybaracolin2018reconfigurable}
Alexei Colin, Emily Ruppel, and Brandon Lucia.
\newblock A reconfigurable energy storage architecture for energy-harvesting
  devices.
\newblock In {\em ACM SIGPLAN Notices}, volume~53, pages 767--781. ACM, 2018.

\bibitem{conti2018xnor}
Francesco Conti, Pasquale~Davide Schiavone, and Luca Benini.
\newblock Xnor neural engine: A hardware accelerator ip for 21.6-fj/op binary
  neural network inference.
\newblock {\em IEEE Transactions on Computer-Aided Design of Integrated
  Circuits and Systems}, 37(11):2940--2951, 2018.

\bibitem{bnn}
Matthieu Courbariaux, Itay Hubara, Daniel Soudry, Ran El-Yaniv, and Yoshua
  Bengio.
\newblock Binarized neural networks: Training deep neural networks with weights
  and activations constrained to+ 1 or-1.
\newblock {\em arXiv preprint arXiv:1602.02830}, 2016.

\bibitem{m2diao2007spin}
Zhitao Diao, Alex Panchula, Yunfei Ding, Mahendra Pakala, Shengyuan Wang,
  Zhanjie Li, Dmytro Apalkov, Hideyasu Nagai, Alexander Driskill-Smith,
  Lien-Chang Wang, et~al.
\newblock Spin transfer switching in dual mgo magnetic tunnel junctions.
\newblock {\em Applied Physics Letters}, 90(13):132508, 2007.

\bibitem{dong2008circuit}
Xiangyu Dong, Xiaoxia Wu, Guangyu Sun, Yuan Xie, Helen Li, and Yiran Chen.
\newblock Circuit and microarchitecture evaluation of 3d stacking magnetic ram
  (mram) as a universal memory replacement.
\newblock In {\em 2008 45th ACM/IEEE Design Automation Conference}, pages
  554--559. IEEE, 2008.

\bibitem{nvsimdong2012nvsim}
Xiangyu Dong, Cong Xu, Yuan Xie, and Norman~P Jouppi.
\newblock Nvsim: A circuit-level performance, energy, and area model for
  emerging nonvolatile memory.
\newblock {\em IEEE Transactions on Computer-Aided Design of Integrated
  Circuits and Systems}, 31(7):994--1007, 2012.

\bibitem{m1durrenfeld2015tunable}
Philipp D{\"u}rrenfeld, Felicitas Gerhard, Jonathan Chico, Randy~K Dumas,
  Mojtaba Ranjbar, Anders Bergman, Lars Bergqvist, Anna Delin, Charles Gould,
  Laurens~W Molenkamp, et~al.
\newblock Tunable damping, saturation magnetization, and exchange stiffness of
  half-heusler nimnsb thin films.
\newblock {\em Physical Review B}, 92(21):214424, 2015.

\bibitem{neuralcacheeckert2018neural}
Charles Eckert, Xiaowei Wang, Jingcheng Wang, Arun Subramaniyan, Ravi Iyer,
  Dennis Sylvester, David Blaauw, and Reetuparna Das.
\newblock Neural cache: Bit-serial in-cache acceleration of deep neural
  networks.
\newblock In {\em Proceedings of the 45th Annual International Symposium on
  Computer Architecture}, pages 383--396. IEEE Press, 2018.

\bibitem{whatsnextganesan2019s}
Karthik Ganesan, Joshua San~Miguel, and Natalie~Enright Jerger.
\newblock The what's next intermittent computing architecture.
\newblock In {\em 2019 IEEE International Symposium on High Performance
  Computer Architecture (HPCA)}, pages 211--223. IEEE, 2019.

\bibitem{dnngobieski2018intermittent}
Graham Gobieski, Nathan Beckmann, and Brandon Lucia.
\newblock Intermittent deep neural network inference, 2018.

\bibitem{InferenceBeyondEdge}
Graham Gobieski, Brandon Lucia, and Nathan Beckmann.
\newblock Intelligence beyond the edge: Inference on intermittent embedded
  systems.
\newblock In {\em Proceedings of the Twenty-Fourth International Conference on
  Architectural Support for Programming Languages and Operating Systems}, pages
  199--213. ACM, 2019.

\bibitem{implantgreenspan2016guest}
Hayit Greenspan, Bram Van~Ginneken, and Ronald~M Summers.
\newblock Guest editorial deep learning in medical imaging: Overview and future
  promise of an exciting new technique.
\newblock {\em IEEE Transactions on Medical Imaging}, 35(5):1153--1159, 2016.

\bibitem{mcu2hester2016amulet}
Josiah Hester, Travis Peters, Tianlong Yun, Ronald Peterson, Joseph Skinner,
  Bhargav Golla, Kevin Storer, Steven Hearndon, Kevin Freeman, Sarah Lord,
  et~al.
\newblock Amulet: An energy-efficient, multi-application wearable platform.
\newblock In {\em Proceedings of the 14th ACM Conference on Embedded Network
  Sensor Systems CD-ROM}, pages 216--229. ACM, 2016.

\bibitem{mcu3hester2015tragedy}
Josiah Hester, Lanny Sitanayah, and Jacob Sorber.
\newblock Tragedy of the coulombs: Federating energy storage for tiny,
  intermittently-powered sensors.
\newblock In {\em Proceedings of the 13th ACM Conference on Embedded Networked
  Sensor Systems}, pages 5--16. ACM, 2015.

\bibitem{mcu4hester2017flicker}
Josiah Hester and Jacob Sorber.
\newblock Flicker: Rapid prototyping for the batteryless internet-of-things.
\newblock In {\em Proceedings of the 15th ACM Conference on Embedded Network
  Sensor Systems}, page~19. ACM, 2017.

\bibitem{mayflyhester2017timely}
Josiah Hester, Kevin Storer, and Jacob Sorber.
\newblock Timely execution on intermittently powered batteryless sensors.
\newblock In {\em Proceedings of the 15th ACM Conference on Embedded Network
  Sensor Systems}, page~17. ACM, 2017.

\bibitem{clankhicks2017clank}
Matthew Hicks.
\newblock Clank: Architectural support for intermittent computation.
\newblock In {\em 2017 ACM/IEEE 44th Annual International Symposium on Computer
  Architecture (ISCA)}, pages 228--240. IEEE, 2017.

\bibitem{m2hu2015stt}
G~Hu, JH~Lee, JJ~Nowak, JZ~Sun, J~Harms, A~Annunziata, S~Brown, W~Chen, YH~Kim,
  G~Lauer, et~al.
\newblock Stt-mram with double magnetic tunnel junctions.
\newblock In {\em 2015 IEEE International Electron Devices Meeting (IEDM)},
  pages 26--3. IEEE, 2015.

\bibitem{jayakumar2014quickrecall}
Hrishikesh Jayakumar, Arnab Raha, and Vijay Raghunathan.
\newblock Quickrecall: A low overhead hw/sw approach for enabling computations
  across power cycles in transiently powered computers.
\newblock In {\em 2014 27th International Conference on VLSI Design and 2014
  13th International Conference on Embedded Systems}, pages 330--335. IEEE,
  2014.

\bibitem{energyharvestingkim2014ambient}
Sangkil Kim, Rushi Vyas, Jo~Bito, Kyriaki Niotaki, Ana Collado, Apostolos
  Georgiadis, and Manos~M Tentzeris.
\newblock Ambient rf energy-harvesting technologies for self-sustainable
  standalone wireless sensor platforms.
\newblock {\em Proceedings of the IEEE}, 102(11):1649--1666, 2014.

\bibitem{ADULTkohavi1996scaling}
Ron Kohavi.
\newblock Scaling up the accuracy of naive-bayes classifiers: A decision-tree
  hybrid.
\newblock In {\em Kdd}, volume~96, pages 202--207. Citeseer, 1996.

\bibitem{MNISTlecun1998gradient}
Yann LeCun, L{\'e}on Bottou, Yoshua Bengio, Patrick Haffner, et~al.
\newblock Gradient-based learning applied to document recognition.
\newblock {\em Proceedings of the IEEE}, 86(11):2278--2324, 1998.

\bibitem{pinatubo}
Shuangchen Li, Cong Xu, Qiaosha Zou, Jishen Zhao, Yu~Lu, and Yuan Xie.
\newblock Pinatubo: A processing-in-memory architecture for bulk bitwise
  operations in emerging non-volatile memories.
\newblock In {\em Proceedings of the 53rd Annual Design Automation Conference},
  page 173. ACM, 2016.

\bibitem{pudiannao}
Daofu Liu, Tianshi Chen, Shaoli Liu, Jinhong Zhou, Shengyuan Zhou, Olivier
  Teman, Xiaobing Feng, Xuehai Zhou, and Yunji Chen.
\newblock Pudiannao: A polyvalent machine learning accelerator.
\newblock In {\em ACM SIGARCH Computer Architecture News}, volume~43, pages
  369--381. ACM, 2015.

\bibitem{liu2016lightweight}
Qingrui Liu and Changhee Jung.
\newblock Lightweight hardware support for transparent consistency-aware
  checkpointing in intermittent energy-harvesting systems.
\newblock In {\em 2016 5th Non-Volatile Memory Systems and Applications
  Symposium (NVMSA)}, pages 1--6. IEEE, 2016.

\bibitem{processor}
Yongpan Liu, Zewei Li, Hehe Li, Yiqun Wang, Xueqing Li, Kaisheng Ma, Shuangchen
  Li, Meng-Fan Chang, Sampson John, Yuan Xie, et~al.
\newblock Ambient energy harvesting nonvolatile processors: from circuit to
  system.
\newblock In {\em Proceedings of the 52nd Annual Design Automation Conference},
  page 150. ACM, 2015.

\bibitem{dinolucia2015simpler}
Brandon Lucia and Benjamin Ransford.
\newblock A simpler, safer programming and execution model for intermittent
  systems.
\newblock In {\em ACM SIGPLAN Notices}, volume~50, pages 575--585. ACM, 2015.

\bibitem{lukosevicius2017using}
Giedrius Lukosevicius, Alberto~Rodriguez Arreola, and Alex~S Weddell.
\newblock Using sleep states to maximize the active time of transient computing
  systems.
\newblock In {\em Proceedings of the Fifth ACM International Workshop on Energy
  Harvesting and Energy-Neutral Sensing Systems}, pages 31--36. ACM, 2017.

\bibitem{ma2017incidental}
Kaisheng Ma, Xueqing Li, Jinyang Li, Yongpan Liu, Yuan Xie, Jack Sampson,
  Mahmut~Taylan Kandemir, and Vijaykrishnan Narayanan.
\newblock Incidental computing on iot nonvolatile processors.
\newblock In {\em 2017 50th Annual IEEE/ACM International Symposium on
  Microarchitecture (MICRO)}, pages 204--218. IEEE, 2017.

\bibitem{considereverycyclema2015architecture}
Kaisheng Ma, Yang Zheng, Shuangchen Li, Karthik Swaminathan, Xueqing Li,
  Yongpan Liu, Jack Sampson, Yuan Xie, and Vijaykrishnan Narayanan.
\newblock Architecture exploration for ambient energy harvesting nonvolatile
  processors.
\newblock In {\em 2015 IEEE 21st International Symposium on High Performance
  Computer Architecture (HPCA)}, pages 526--537. IEEE, 2015.

\bibitem{alpacamaeng2017alpaca}
Kiwan Maeng, Alexei Colin, and Brandon Lucia.
\newblock Alpaca: intermittent execution without checkpoints.
\newblock {\em Proceedings of the ACM on Programming Languages}, 1(OOPSLA):96,
  2017.

\bibitem{chinchillamaeng2018adaptive}
Kiwan Maeng and Brandon Lucia.
\newblock Adaptive dynamic checkpointing for safe efficient intermittent
  computing.
\newblock In {\em 13th $\{$USENIX$\}$ Symposium on Operating Systems Design and
  Implementation ($\{$OSDI$\}$ 18)}, pages 129--144, 2018.

\bibitem{sensornetworkmanic2016intelligent}
Milos Manic, Kasun Amarasinghe, Juan~J Rodriguez-Andina, and Craig Rieger.
\newblock Intelligent buildings of the future: Cyberaware, deep learning
  powered, and human interacting.
\newblock {\em IEEE Industrial Electronics Magazine}, 10(4):32--49, 2016.

\bibitem{m1mizukami2009low}
S~Mizukami, D~Watanabe, M~Oogane, Y~Ando, Y~Miura, M~Shirai, and T~Miyazaki.
\newblock Low damping constant for co 2 feal heusler alloy films and its
  correlation with density of states.
\newblock {\em Journal of Applied Physics}, 105(7):07D306, 2009.

\bibitem{R}
{R Core Team}.
\newblock {\em R: A Language and Environment for Statistical Computing}.
\newblock R Foundation for Statistical Computing, Vienna, Austria, 2016.

\bibitem{mementosransford2011mementos}
Benjamin Ransford, Jacob Sorber, and Kevin Fu.
\newblock Mementos: system support for long-running computation on rfid-scale
  devices.
\newblock In {\em ACM SIGARCH Computer Architecture News}, volume~39, pages
  159--170. ACM, 2011.

\bibitem{coatiruppel2019transactional}
Emily Ruppel and Brandon Lucia.
\newblock Transactional concurrency control for intermittent, energy-harvesting
  computing systems.
\newblock In {\em Proceedings of the 40th ACM SIGPLAN Conference on Programming
  Language Design and Implementation}, pages 1085--1100. ACM, 2019.

\bibitem{mtjsaida2016sub}
Daisuke Saida, Saori Kashiwada, Megumi Yakabe, Tadaomi Daibou, Naoki Hase,
  Miyoshi Fukumoto, Shinji Miwa, Yoshishige Suzuki, Hiroki Noguchi, Shinobu
  Fujita, et~al.
\newblock Sub-3 ns pulse with sub-100 $\mu$a switching of 1x--2x nm
  perpendicular mtj for high-performance embedded stt-mram towards sub-20 nm
  cmos.
\newblock In {\em 2016 IEEE Symposium on VLSI Technology}, pages 1--2. IEEE,
  2016.

\bibitem{modernMTJsaida2016sub}
Daisuke Saida, Saori Kashiwada, Megumi Yakabe, Tadaomi Daibou, Naoki Hase,
  Miyoshi Fukumoto, Shinji Miwa, Yoshishige Suzuki, Hiroki Noguchi, Shinobu
  Fujita, et~al.
\newblock Sub-3 ns pulse with sub-100 $\mu$a switching of 1x--2x nm
  perpendicular mtj for high-performance embedded stt-mram towards sub-20 nm
  cmos.
\newblock In {\em 2016 IEEE Symposium on VLSI Technology}, pages 1--2. IEEE,
  2016.

\bibitem{mcu5sample2008design}
Alanson~P Sample, Daniel~J Yeager, Pauline~S Powledge, Alexander~V Mamishev,
  and Joshua~R Smith.
\newblock Design of an rfid-based battery-free programmable sensing platform.
\newblock {\em IEEE transactions on instrumentation and measurement},
  57(11):2608--2615, 2008.

\bibitem{ehmodelsan2018eh}
Joshua San~Miguel, Karthik Ganesan, Mario Badr, Chunqiu Xia, Rose Li, Hsuan
  Hsiao, and Natalie~Enright Jerger.
\newblock The eh model: Early design space exploration of intermittent
  processor architectures.
\newblock In {\em 2018 51st Annual IEEE/ACM International Symposium on
  Microarchitecture (MICRO)}, pages 600--612. IEEE, 2018.

\bibitem{m1sato2014properties}
H~Sato, ECI Enobio, M~Yamanouchi, S~Ikeda, S~Fukami, S~Kanai, F~Matsukura, and
  H~Ohno.
\newblock Properties of magnetic tunnel junctions with a mgo/cofeb/ta/cofeb/mgo
  recording structure down to junction diameter of 11 nm.
\newblock {\em Applied Physics Letters}, 105(6):062403, 2014.

\bibitem{ambit}
Vivek Seshadri, Donghyuk Lee, Thomas Mullins, Hasan Hassan, Amirali Boroumand,
  Jeremie Kim, Michael~A Kozuch, Onur Mutlu, Phillip~B Gibbons, and Todd~C
  Mowry.
\newblock Ambit: In-memory accelerator for bulk bitwise operations using
  commodity dram technology.
\newblock In {\em Proceedings of the 50th Annual IEEE/ACM International
  Symposium on Microarchitecture}, pages 273--287. ACM, 2017.

\bibitem{RRAMsu2017462gops}
Fang Su, Wei-Hao Chen, Lixue Xia, Chieh-Pu Lo, Tianqi Tang, Zhibo Wang,
  Kuo-Hsiang Hsu, Ming Cheng, Jun-Yi Li, Yuan Xie, et~al.
\newblock A 462gops/j rram-based nonvolatile intelligent processor for energy
  harvesting ioe system featuring nonvolatile logics and processing-in-memory.
\newblock In {\em 2017 Symposium on VLSI Technology}, pages T260--T261. IEEE,
  2017.

\bibitem{rram2018}
Xiaoyu Sun, Xiaochen Peng, Pai-Yu Chen, Rui Liu, Jae-sun Seo, and Shimeng Yu.
\newblock Fully parallel rram synaptic array for implementing binary neural
  network with (+ 1,- 1) weights and (+ 1, 0) neurons.
\newblock In {\em Proceedings of the 23rd Asia and South Pacific Design
  Automation Conference}, pages 574--579. IEEE Press, 2018.

\bibitem{rram2017}
Tianqi Tang, Lixue Xia, Boxun Li, Yu~Wang, and Huazhong Yang.
\newblock Binary convolutional neural network on rram.
\newblock In {\em Design Automation Conference (ASP-DAC), 2017 22nd Asia and
  South Pacific}, pages 782--787. IEEE, 2017.

\bibitem{HARweb}
https://archive.ics.uci.edu/ml/datasets/human+activity+recognition+using+smartphones,
  2019.
\newblock Accessed: 2019-06-02.

\bibitem{ratchetvan2016intermittent}
Joel Van Der~Woude and Matthew Hicks.
\newblock Intermittent computation without hardware support or programmer
  intervention.
\newblock In {\em 12th $\{$USENIX$\}$ Symposium on Operating Systems Design and
  Implementation ($\{$OSDI$\}$ 16)}, pages 17--32, 2016.

\bibitem{wang2016magnetic}
Jian-Ping Wang, Mahdi Jamaliz, Angeline~Klemm Smith, and Zhengyang Zhao.
\newblock Magnetic tunnel junction based integrated logics and computational
  circuits.
\newblock {\em Nanomagnetic and Spintronic Devices for Energy-Efficient Memory
  and Computing}, page 133, 2016.

\bibitem{rram2016b}
Lixue Xia, Tianqi Tang, Wenqin Huangfu, Ming Cheng, Xiling Yin, Boxun Li,
  Yu~Wang, and Huazhong Yang.
\newblock Switched by input: power efficient structure for rram-based
  convolutional neural network.
\newblock In {\em Proceedings of the 53rd Annual Design Automation Conference},
  page 125. ACM, 2016.

\bibitem{rram2016}
Shimeng Yu, Zhiwei Li, Pai-Yu Chen, Huaqiang Wu, Bin Gao, Deli Wang, Wei Wu,
  and He~Qian.
\newblock Binary neural network with 16 mb rram macro chip for classification
  and online training.
\newblock In {\em Electron Devices Meeting (IEDM), 2016 IEEE International},
  pages 16--2. IEEE, 2016.

\bibitem{she}
Masoud Zabihi, Zhengyang Zhao, DC~Mahendra, Zamshed~I Chowdhury, Salonik Resch,
  Thomas Peterson, Ulya~R Karpuzcu, Jian-Ping Wang, and Sachin~S Sapatnekar.
\newblock Using spin-hall mtjs to build an energy-efficient in-memory
  computation platform.
\newblock In {\em 20th International Symposium on Quality Electronic Design
  (ISQED)}, pages 52--57. IEEE, 2019.

\end{thebibliography}

\end{document}